# Synthesis, floating zone crystal growth and characterization of the Quantum Spin Ice $Pr_2Zr_2O_7$ pyrochlore

S.M. Koohpayeh[a], J.-J. Wen[a], B.A. Trump[a,b], C.L. Broholm[a], and T.M. McQueen[a,b]

[a] Institute for Quantum Matter and Department of Physics and Astronomy, Johns Hopkins University, Baltimore, MD 21218, USA

[b] Department of Chemistry, Johns Hopkins University, Baltimore, MD 21218, USA

Corresponding author: Dr S.M. Koohpayeh, Koohpayeh@jhu.edu
Tel. +1 410 516 7687; Fax +1 410 516 7239

**Abstract**

Pyrochlore $Pr^{3+}_{2+x}Zr^{4+}_{2-x}O_{7-x/2}$ samples in the form of both powders ($-0.02 \leq x \leq 0.02$) and bulk single crystals have been studied to elucidate the dependence of their magnetic, compositional and structural properties on synthesis and growth conditions. All samples were characterized using X-ray diffraction, specific heat, and DC magnetization measurements. The crystals were also studied using the X-ray Laue technique and scanning electron microscopy. Increasing the Pr content for the $Pr_{2+x}Zr_{2-x}O_{7-x/2}$ powders enlarged the lattice parameter, and resulted in systematic changes in magnetic susceptibility and specific heat. Stoichiometric and high quality single crystals of $Pr_2Zr_2O_7$ were grown using the optical floating zone technique under a high purity static argon atmosphere, to avoid inclusions of $Pr^{4+}$ and limit Pr vaporization. Increasing the growth speed was found to significantly reduce Pr vaporization for better control of stoichiometry. Scanning electron microscopy provided direct evidence of spinodal decomposition during growth that is controllable via rotation rate. An intermediate rotation rate of 3-6 rpm was found to produce the best microstructure. The magnetic susceptibility of crystals grown at rates from 1-20 mm/hr revealed changes that were consistent with Pr vaporization. Further, we report indications of local off-centering of $Pr^{3+}$ ions from the ideal pyrochlore sites, similar to what is known for the trivalent cation in $Bi_2Ti_2O_7$ and $La_2Zr_2O_7$. The effect varies with Pr content and radically modulates the low temperature specific heat. Overall, the results clearly demonstrate important correlations between the growth conditions and physical properties of $Pr_2Zr_2O_7$ crystals.

## 1. Introduction

Pyrochlore oxides have been intensively studied experimentally and theoretically for the last few decades in search of novel magnetism associated with its frustrated magnetic lattice [1-6]. Cubic pyrochlores $A_2B_2O_7$ belong to the space group $Fd\bar{3}m$, where the A site is occupied by a trivalent cation with eightfold oxygen coordination and the B site is occupied by a tetravalent transition metal ion with sixfold oxygen coordination [7].

Previously pyrochlore oxides have been synthesized and prepared as polycrystalline samples [8-12], small single crystals (mm size) by flux [13-15], and vapor transport [16] methods. More recently larger single crystals of $R_2Ti_2O_7$ (R=rare earth) [17-21] and $Pr_2Zr_2O_7$ [22-24] have been grown using the optical floating zone furnace. The advantage of this technique is that the use of crucibles (in particular, for the growth of such high melting point oxides for which crucibles are undesirable) and pick up of impurities (a common problem with flux and Czochralski growth techniques) are avoided [25]. However, despite growing larger crystals of reported congruently melting compounds of $R_2Ti_2O_7$ using an optical floating zone furnace, the magnetic properties of the resulting crystals (e.g. $Tb_2Ti_2O_7$) have been shown to vary significantly from one sample to another [6]. For $Pr_2Zr_2O_7$, a starting composition enriched with Pr is reported to help keep the Pr/Zr ratio to 1 when the growth is performed in an oxygen atmosphere [22-23].

Compounding the complexities of the single crystal growth process and the thermodynamics involved, the synthesis and growth of $A_2B_2O_7$ oxides are further complicated by the different cation oxidation states and lattice disordering tendencies that are known to occur in the pyrochlore structure [26-30]. While the ionic size, charge and electronic effects are determining factors on these tendencies, the pyrochlore can be transformed into the disordered fluorite structure (space group $Fm\bar{3}m$) by mixing the A and B atoms (on cation sublattice) and randomly distributing the oxygen vacancies (on the anion sublattice). To better understand the stability of the pyrochlore structure, more detailed theoretical and experimental studies have been reported on the defect formation energy, oxygen positional parameter, and disorder mechanisms of cation antisite and oxygen Frenkel defects [26, 31-32]. Furthermore, a recent study of order-disorder transformation temperature based on temperature-composition phase diagrams for various $ZrO_2$-$A_2O_3$ systems has shown that the pyrochlore structure of $A_2Zr_2O_7$ for the largest ionic radius of A (such as La and Pr) is more stable at high temperatures close to the melting points [33].

However, little if any, of the published work on the crystal growth of $A_2B_2O_7$ appears to have been investigated in detail to determine whether the growth parameters have any effect on crystal quality. Through a systematic study, the aim of the present work was therefore to prepare and characterize both polycrystalline $Pr_2Zr_2O_7$ (by a standard solid state process) and high quality single crystal samples (using the optical floating zone technique), and to discover what effects varying preparation conditions have upon the quality, structural, and magnetic properties of the crystals prepared.

In polycrystalline studies, we find substantial changes in low temperature specific heat from small changes in Pr:Zr ratio, likely due to an off-centering of the Pr ions. In single crystal studies, we find vaporization of Pr, incorporation of $Pr^{4+}$ in the crystal, and spinodal decomposition during floating zone growth, which can be greatly suppressed by use of an inert argon atmosphere and optimization of crystal rotation and growth rate. An optimized set of optical floating zone crystal growth parameters that produce single crystals with structural and physical properties virtually identical to those of the stoichiometric polycrystalline reference are reported.

## 2. Experimental procedures

### 2.1. Synthesis of polycrystalline $Pr_{2+x}Zr_{2-x}O_{7-x/2}$ (-0.02≤x≤0.02)

Polycrystalline samples of $Pr_{2+x}Zr_{2-x}O_{7-x/2}$ with different Pr content (-0.02≤x≤0.02) were prepared to study the effects of Pr non-stoichiometry on the structural and physical properties. These samples of known stoichiometry were then used to tune the composition and quality of as-grown single crystals, due to the vaporization of Pr at temperatures close to melting point of $Pr_2Zr_2O_7$ during the crystal growth process.

To prepare $Pr_{2+x}Zr_{2-x}O_{7-x/2}$ powders the starting materials of $Pr_6O_{11}$ (99.99%, Alfa Aesar) and $ZrO_2$ (99.978%, Alfa Aesar) were thoroughly ground, then heated several times at temperatures of 1350 °C, 1500 °C (as pellet), and 1550 °C (as pellet) in air for 8 hr (at each temperature) with an intermediate grinding. The powder was then sealed in a rubber tube, evacuated, and compacted into a rod (typically 5 mm in diameter and 80 mm long) using a hydraulic press under an isostatic pressure of 70 MPa. After removal from the rubber tube, the rods were sintered at 1550 °C for 10 hr in air. These sintered rods were light brownish due to the presence of some $Pr^{4+}$ that results from heating in an oxygen containing atmosphere. (The incorporation of additional oxygen ions introduces extra negative charge, which is accommodated by increasing the cation charge from $Pr^{3+}$ to $Pr^{4+}$.) Such color change from the black color of $Pr_6O_{11}$ (having both $Pr^{3+}$ and $Pr^{4+}$ ions) to a brownish color for $Pr_5O_9$ and $Pr_7O_{12}$, and then to green color for $Pr_2O_3$ (having only $Pr^{3+}$) has been reported for the praseodymium-oxygen system upon increasing the temperature and decreasing the oxygen content [34].

Therefore, the feed rods were sintered once more at a higher temperature (above 1550 °C and less than the melting temperature of $Pr_2Zr_2O_7$, ~2300 °C) in a four-mirror optical floating zone furnace (Crystal Systems Inc. FZ-T-4000-H-VII-VPO-PC with 4×1 kW halogen lamps) using 80% of the lamp power and a zoning rate of 6 mm/hr under 1 bar static high purity argon atmosphere. No vaporization was detected during this sintering/zoning process. This produced more dense polycrystalline rods with the green color for all the $Pr_{2+x}Zr_{2-x}O_{7-x/2}$ (-0.02≤$x$≤+0.02) samples with no apparent difference in color. Samples for powder X-ray diffraction and lattice parameter measurements, heat capacity and magnetic characterization were cut from these rods.

### 2.2. Single crystal growth

Single crystals of approximately 4 mm in diameter and up to 80 mm in length were grown from the polycrystalline feed rods in a four-mirror optical floating zone furnace (Crystal Systems Inc. FZ-T-12000-X-VPO-PC) with 4 ×3 kW xenon lamps. During all of the growths, the molten zone was moved upwards with the seed crystal being at the bottom and the feed rod above it. This was accomplished by holding the mirrors in a fixed position and translating both the seed and feed rods downwards. Growths were carried out under 1 bar static high-purity argon atmosphere at rates of travel between 1 and 20 mm/hr. Rotation rates of 0 to 12 rpm were employed for the growing crystal (lower shaft) and the feed rod (upper shaft) in opposite directions. In all runs, only one zone pass was performed.

### 2.3. Characterization

Powder X-ray diffraction (XRD) patterns were obtained using a Bruker D8 Focus X-ray diffractometer operating with Cu $K\alpha$ ($K_{\alpha 1}$ = 1.5406 Å) radiation and a Lynxeye silicon strip detector. Phase identification and unit cell determinations were carried out using the Bruker TOPAS software (Bruker AXS). Synchrotron X-ray diffraction data was collected on the high resolution 11-BM-B diffractometer at the Advanced Photon Source, Argonne National Laboratory, with an incident wavelength of $\lambda$ = 0.41369 Å. Silicon was used as an internal standard both for synchrotron and powder X-ray diffraction. All reported uncertainties reflect the statistical portion of the error only. Back-reflection X-ray Laue diffraction (with the X-ray beam of about 1 mm in diameter) was utilized to check the orientation and crystalline quality of the crystals. The compositional homogeneity of polished surface of samples cut directly from the cross sections of the as-grown crystals was probed using a JEOL 6700F scanning electron microscope (SEM) operating in backscatter mode. Magnetic susceptibility measurements were carried out using a Quantum Design Physical Properties Measurement System (PPMS). Temperature-dependent magnetization measurements were collected under applied fields of $\mu_0 H$ = 0.1 T (between $T$ = 2 and 150 K) and 0.5 T (for temperatures above $T$ = 150 K) after zero field cooling (ZFC). The temperature dependence of the specific heat was measured above $T$ = 2 K using the semiadiabatic pulse technique.

## 3. Results and discussion

### 3.1. Characterization of polycrystalline $Pr_{2+x}Zr_{2-x}O_{7-x/2}$

Powder X-ray diffraction confirmed the cubic pyrochlore structure for all of the samples prepared with no detectable evidence of secondary phases. Lattice parameter measurements of the polycrystalline samples prepared with different compositions (-0.02≤$x$≤+0.02) showed larger lattice parameters for higher Pr content, as shown in Fig. 1. The room temperature lattice parameter of $a$ = 10.70254(4) Å corresponds to the stoichiometric $Pr_2Zr_2O_7$ reference sample.

Fig. 2a shows the inverse magnetic susceptibility, estimated by *H/M*, as a function of temperature for finely ground powder samples. The inset in Fig. 2a shows the systematic changes with Pr content. With increasing Pr content the effective magnetic moment (related to the inverse slope) increases, and the Weiss temperature (related to the intercept) becomes less negative. The specific heat of all samples in the form of thin plates (cut from the sintered rods) was measured and is shown in Fig. 2b. The formation of a broad peak *T* ~ 2 K is indicative of short range magnetic correlation and is observed in all samples. *Cp* at temperatures *T* ~ 10 K decreases systematically as the Pr concentration increases. Considering the magnetic interaction energy scale indicated for example by the Curie-Weiss temperature of *T* ~ 1 K [24], changes in *Cp* at 10 K are likely to be associated with changes in single ion properties, such as crystal field levels.

Such large changes in a bulk property like the specific heat are unexpected given the very small change in material composition (0.98≤Pr/Zr≤1.02). To investigate this in greater detail, high-resolution synchrotron X-ray diffraction data were collected on three samples spanning the composition range. Rietveld refinements are shown in Fig. 3a-c, and full refinement parameters are given in Table I. The inclusion of anisotropic displacement parameters is validated at the 99% probability level by the Hamilton R ratio test ($R_{0.01,\ 2,\ 346}$ = 1.0134, $R(F^2)_{iso}/R(F^2)_{aniso}$ = 1.0195 - 1.0576) [39]. A systematic statistical analysis [38] was used to test for the following types of structural imperfections: deviations of Pr:Zr ratio from the initial composition, anti-site mixing, excess oxygen in the ideally vacant O" site located in the center of $Zr_4$ tetrahedra, and oxygen vacancies in the primary lattice. Within the precision of the measurements, we find no further deviations from the ideal pyrochlore structure and formula except the small mixing of Pr on the B site in Pr excess samples (as expected) and Zr mixing on the A site in Zr excess samples (also as expected).

Additional Pr/Zr mixing, beyond that expected for stoichiometry reasons, could be the origin of the large changes in the low temperature specific heat. To test the sensitivity of the refinements to even small amounts of Pr/Zr mixing, the (311) pyrochlore supercell reflection for stoichiometric $Pr_2Zr_2O_7$ is shown in Fig. 3d. The inclusion of Pr on the B site and Zr on the A site (Pr/Zr mixing) visibly worsens the match of the predicted intensity to the data. Further, the changes in the intensities of the (311) pyrochlore reflection for all three samples, shown in Fig. 3e, are well described by models including only the changes in stoichiometry with no additional Pr/Zr mixing. This implies that Pr/Zr mixing is not responsible for the large changes in the low temperature specific heat. Detailed analysis of the Pr anisotropic atomic displacement parameters suggests a possible alternate explanation. The $Pr^{3+}$ ions have a 'pancake'-like atomic displacement, which along with the average size of the atomic displacement parameter for the O' site (within the $Pr_4$ tetrahedra), changes systematically with Pr content: increasing Pr content, reduces the pancake-like nature of the Pr ellipsoids (Fig. 3f) and increases the O' $U_{iso}$ parameter. The pancake like ellipsoids of Pr are similar to that found for Bi in the bismuth pyrochlores $Bi_2Ti_2O_7$ and $Bi_2Ru_2O_7$, which have significant off-centering of the $Bi^{3+}$ ions due to a lone pair effect [40-41]. This suggests $Pr^{3+}$ ions in $Pr_2Zr_2O_7$ do not lie on the ideal crystallographic position although the deviation are too small to produce a stable refinement when displaced off-center. Previous work on $La_2Zr_2O_7$ suggests there are accessible low energy crystabolite-like deformations of the $La_4O'$ sublattice, which would produce such off-centering [42-43], and it is natural to expect similar behavior in $Pr_2Zr_2O_7$. Whether Pr off-centering is static or dynamic in nature, these results would explain why there are large variations in the low temperature specific heat due to small changes in Pr:Zr ratio. Local structure studies are necessary to determine the atomistic details (especially the changes in the O' atomic displacement parameter) and test for very small levels of fluorite-type Pr/Zr mixing, but, regardless of the precise nanoscopic origin, these results imply that all studies must carefully consider and account for local structural complexities.

### 3.2. Floating zone growth and characterization of crystals

Initial floating zone crystal growth attempts using $Pr_2Zr_2O_7$ ($x$ = 0) feed rods in oxygen atmospheres led to significant vaporization of Pr (in the form of Pr oxides), and as a result the grown crystals were non-stoichiometric and dark brown in color, indicating $Pr^{4+}$ inclusions. Successful crystal growths were performed under 1 bar static high-purity argon atmosphere, which significantly helped to control the vaporization and obtain more stoichiometric green crystals (as expected for $Pr^{3+}$). A typical $Pr_2Zr_2O_7$ crystal, grown in argon, is shown in Fig. 4. X-ray Laue patterns taken at regular intervals along the lengths and cross sections of the crystals grown under argon indicated homogeneous high crystalline quality with no detectable variation of orientation and no evidence of spot splitting or distortion.

To investigate whether the rotation rate of the growing crystal and feed rod (as one of the important experimental parameters during floating zone crystal growth [25]) has any effect on crystal quality, crystals were grown at rotation rates of 0, 3, 6 and 12 rpm at a constant growth rate of 4 mm/hr. Although this growth was performed under high purity argon and the crystals were green in color, a small amount of Pr vaporization in the form of Pr oxides was initially observed and confirmed by X-ray diffraction. As will be explained in detail later, the use of a relatively slow growth rate of 4 mm/hr was found to increase the overall amount of vaporization.

Fig. 5 shows back-scattered SEM micrographs taken from polished cross sections of crystals grown at different rotation rates. The micrograph shown for the crystal grown without rotation (Fig. 5a) for which efficient mixing (in the molten zone, in particular, at the liquid/solid interface) and homogeneous heating (within the melt and around the growing crystal) were not achieved, indicative of spinodal decomposition. While SEM/EDX (energy dispersive X-ray spectroscopy) analysis is not expected to provide the exact quantitative composition of the two domains, the back-scattered images clearly show brighter areas with more of the heavier element Pr, while the areas with the lighter element Zr appear darker due to less backscattered electrons. Such phase decomposition at high temperatures is not reported in the calculated phase diagram of $ZrO_2$ – $PrO_{1.5}$ [35], however, the experimentally illustrated phase diagram of O-Pr-Zr system [36] predicts fluorite-type phases very close to the composition and melting temperature of $Pr_2Zr_2O_7$. More concentration of the Pr rich areas close to the edge of cross section (near the surface of the crystal) can be seen on the smaller picture shown on the top right of Fig. 5a. This is consistent with the tendency of Pr to vaporize in this system as observed during the crystal growth process.

Moreover, as seen in Fig 5b-c, applying the slow rotation rate of 3 rpm and 6 rpm yielded a significantly more uniform microstructure with no obvious phase separation or inclusions on the micro-meter length scale. Although internal compositional homogeneity was achieved at the faster rotation rate of 12 rpm, it appears that some Pr rich phases were also pushed to the surface. Based on the fact that rotation has a complicated dynamic nature (involving heat transformation, motion and force in the region near to the liquid/solid interface where the crystal is grown [25]), it can be concluded that the slower rotation rates of 3 and 6 rpm provides efficient mixing, heating, and force in the molten zone that led to the formation of a more homogeneous microstructure. However, using a higher rotation rate of 12 rpm (which enhances forced convection in the molten zone) affects the phase distribution at the liquid/solid interface in such a way that some of Pr rich phases move toward the edge or surface of the crystal (as shown in the top right of Fig. 5d).

All the previous growth attempts using a slower growth rate of 3-4 mm/hr with a rotation rate of 12 rpm produced similar quality crystals, which developed light colored powders on the crystal surfaces when exposed to air for an extended period of time under ambient conditions. XRD analysis of this light powder collected from the surface indicated the pyrochlore $Pr_2Zr_2O_7$ and a fluorite type structure of $PrO_2$, which should have been formed at temperatures close to the melting point of $Pr_2Zr_2O_7$ (as closely predicted in the phase diagram of the O-Pr-Zr system [36]). These pyrochlore and fluorite structures are closely related and have been discussed in detail in earlier reports [26, 31-33].

XRD of crushed single crystals grown at different rotation rates showed similar pyrochlore reflections with small changes in the lattice parameter and no obvious inclusions. The lattice parameter measured for the crystal grown at 0 rpm was $a$ = 10.68613(4) Å, while it decreased to $a$ = 10.68465(4) Å and $a$ = 10.68160(4) Å for rotation rates of 6 and 12 rpm respectively. These data indicate that Pr vaporization is enhanced by increasing the rotation rate, because this increases mixing and transports more of the molten material to the surface where vaporization can then occur.

Looking more carefully at the peak shapes of XRD patterns also revealed that the crystal grown without rotation has broader diffraction peaks. Increased rotation rate during growth decreases XRD peak broadening and peaks become more similar to those of the stoichiometric polycrystalline $Pr_2Zr_2O_7$ reference sample (see Figs. 6a and 6b). Considering the same instrumental broadening, sample contribution to peak broadening can arise from crystal size and/or some form of strain (micro strain due to dislocations, concentration gradients, and structural defects such as stacking faults or twinning.) [44]. Changes in stoichiometry due to the different rotation rates alone seems to be an unlikely factor in peak broadening because the more non-stoichiometric crystals (grown at 6 and 12 rpm with smaller lattice parameters) showed even sharper peaks. It should be noted that the double peaks seen in Fig. 6b are due to an instrumental effect (the $K\alpha_1$ and $K\alpha_2$ lines from the X-ray source).

By using a Williamson-Hall analysis [45] the peak broadening effects due to size and strain can be separated and estimated. The resulting Williamson-Hall plot is shown in Fig. 6c (parameters in Table II) with linear fits of the form: $\beta_{obs}\cos\theta = C\varepsilon\sin\theta + K\lambda/L$. Here C is a constant, ε represents apparent strain, K is a constant close to unity, λ is the incident wavelength, and L is the average crystallite size. Though several assumptions are made in the Williamson-Hall analysis, the slope of each line indicates that the slower rotation rates (0 rpm and 3 rpm) produce more strain than faster rotation rates (6 rpm and 12 rpm). The highest strain is observed for 0 rpm, and reflects the spinodal decomposition observed by SEM. This observation also agrees with the data in Fig. 6a and 6b, as the 0 rpm and 3 rpm samples have broader peak tails, while the 6 rpm and 12 rpm samples line up with the stoichiometric powder. Consistent with the SEM data in Fig. 5, the Williamson-Hall plot indicates the sample grown with a rotation rate of 6 rpm has the least strain and is the best sample from the perspective of microstructure.

As one of the other critical growth parameters governing crystal quality [25], the effects of growth rate (speed) was also examined from the slow growth rate of 1 mm/hr to the faster rate of 20 mm/hr. As shown in Fig. 7, it was found that the faster growth rates considerably reduced vaporization of Pr and yielded more stoichiometric crystals, with $a$ = 10.70088(2) Å (at 20 mm/hr) close to the pure, stoichiometric $Pr_2Zr_2O_7$ powder ($a$ = 10.70254(4) Å).

Although increasing the growth rate to 20 mm/hr significantly reduced vaporization, crack formation due to the higher cooling rate was then found to become an issue [37]. Therefore, to grow stoichiometric, crack free, homogeneous single crystals, an intermediate growth speed of 10 mm/hr was applied, an extra 1 atomic% Pr was added into the starting feed rods (to offset Pr vaporization), and a rotation rate of 6 rpm was used. These optimized growth conditions lead to high quality crystals with the lattice parameter of $a$ = 10.70227(5) Å (also shown in Fig. 7) very similar to the pure $Pr_2Zr_2O_7$ reference powder. We note that the lattice parameter of these stoichiometric crystals are significantly closer to the ideal $Pr_2Zr_2O_7$ (determined from polycrystalline syntheses where the compositions are known) than previous reported growths under oxygen atmospheres [22-23]. The optimized grown crystals have been stable for months in ambient conditions, with no surface degradation.

Magnetic susceptibility and heat capacity measurements were also performed on the single crystals grown at different growth speeds and on the crystal grown under optimized conditions, as shown in Fig. 8. The observed trend is consistent with Fig. 2 and this confirms that growth speed is a key factor in controlling the Pr concentration of the grown crystals. A Curie-Weiss fit of the data between 2 K and 20 K (Fig. 9) also reveals a larger effective moment with higher Pr content (and larger lattice parameter) and proximity to the reference sample for crystals grown under the optimized conditions. Full single crystal X-ray diffraction data for the optimized specimen are reported elsewhere [46]. The optimized grown crystal has physical properties virtually identical to that of the stoichiometric powder sample, attesting to its quality.

## 4. Conclusions

The stoichiometry and homogeneity of $Pr_2Zr_2O_7$ are strongly influenced by growth conditions and in turn significantly affect the structural and physical properties. The use of high purity argon gas during sintering and floating zone crystal growth of $Pr_2Zr_2O_7$ is crucial to considerably reduce Pr vaporization and prevent the formation of $Pr^{4+}$ ions. The rotation rate and growth speed were found to be parameters with major influence upon the eventual quality of $Pr_2Zr_2O_7$ single crystals due to phase homogeneity and vaporization respectively. Increasing the rotation rate leads to the formation of a more homogeneous microstructure while increasing the growth speed limits Pr vaporization and single crystals with lattice parameters closer to the stoichiometric $Pr_2Zr_2O_7$ reference powder were obtained. Pr deficient samples had smaller lattice parameters, and systematic changes were observed in magnetic susceptibility and heat capacity data dependent on the Pr content. There is likely a local off-centering of $Pr^{3+}$ ions from the ideal pyrochlore sites in stoichiometric samples, similar to what is known to occur in $Bi_2Ti_2O_7$ and $La_2Zr_2O_7$. This effect inversely depends on Pr content at the atomic % level and may be the cause of the radically modulated low temperature specific heat. These results clearly demonstrate important correlations between growth conditions and physical properties of $Pr_2Zr_2O_7$ crystals, and have allowed for the development of an optimized growth procedure that for the first time produces crystals with structural and physical properties matching those of the stoichiometric polycrystalline reference standard. The structural complexity in this 'simple' pyrochlore system likely apply to many related pyrochlores of current interest in the condensed matter physics community and must be accounted for to achieve understanding of the physical properties.

**Acknowledgments**

This work was supported by U.S. Department of Energy (DOE), Office of Basic Energy Sciences, Division of Materials Sciences and Engineering under award DE-FG02-08ER46544. Use of the Advanced Photon Source at Argonne National Laboratory was supported by the U. S. Department of Energy, Office of Science, Office of Basic Energy Sciences, under contract No. DE-AC02-06CH11357. Fruitful discussions with A. M. Fry, Satoru Nakatsuji and Kenta Kimura are gratefully acknowledged.

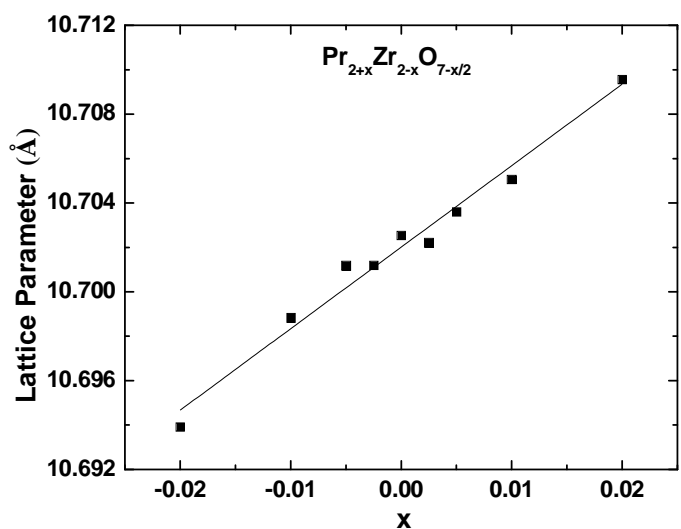

**Fig. 1.** Lattice parameters of $Pr_{2+x}Zr_{2-x}O_{7-x/2}$ at different $x$ (-0.02≤$x$≤0.02) measured by powder X-ray diffraction at room temperature using the Bruker TOPAS software (Bruker AXS). Samples with more Pr content showed larger lattice parameters. A lattice parameter of $a$ = 10.70254(4) Å was measured for the stoichiometric $Pr_2Zr_2O_7$ sample. All quoted error bars for lattice parameters includes only the statistical contribution to the error. A best fit line is shown.

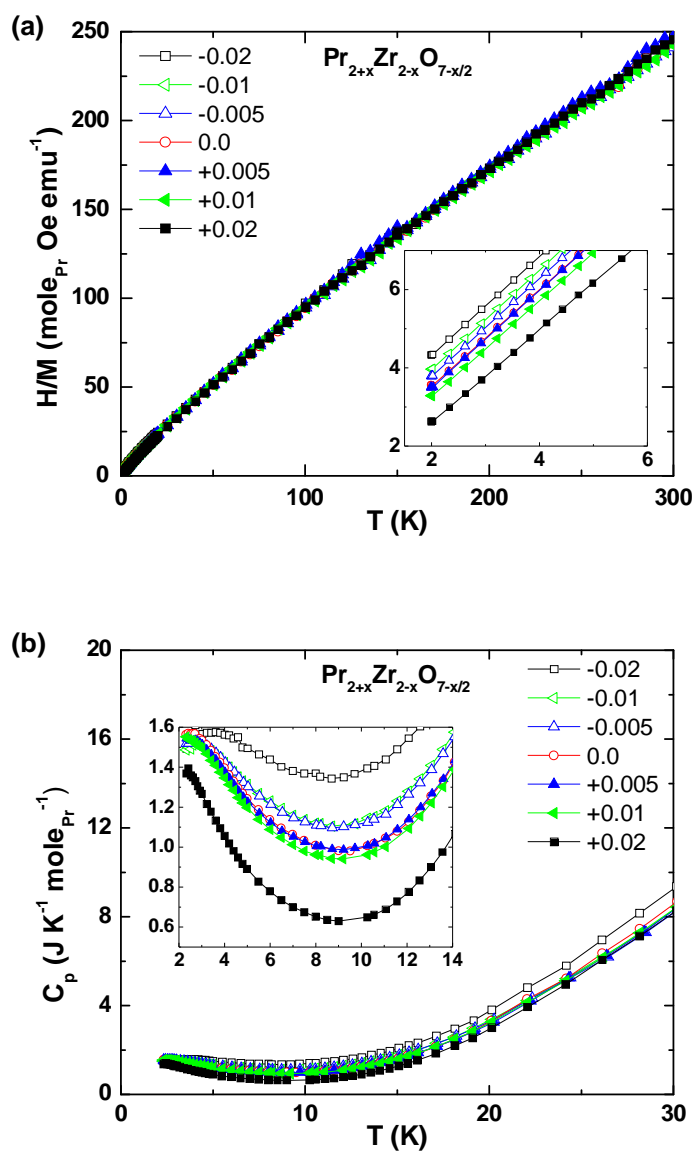

**Fig. 2.** Temperature dependence of inverse magnetic susceptibility (a) and specific heat (b) of polycrystalline $Pr_{2+x}Zr_{2-x}O_{7-x/2}$ samples. Insets are shown for each to demonstrate the behavior at low temperatures. Systematic changes across the series show strong correlation between physical properties and Pr concentration in $Pr_{2+x}Zr_{2-x}O_{7-x/2}$.

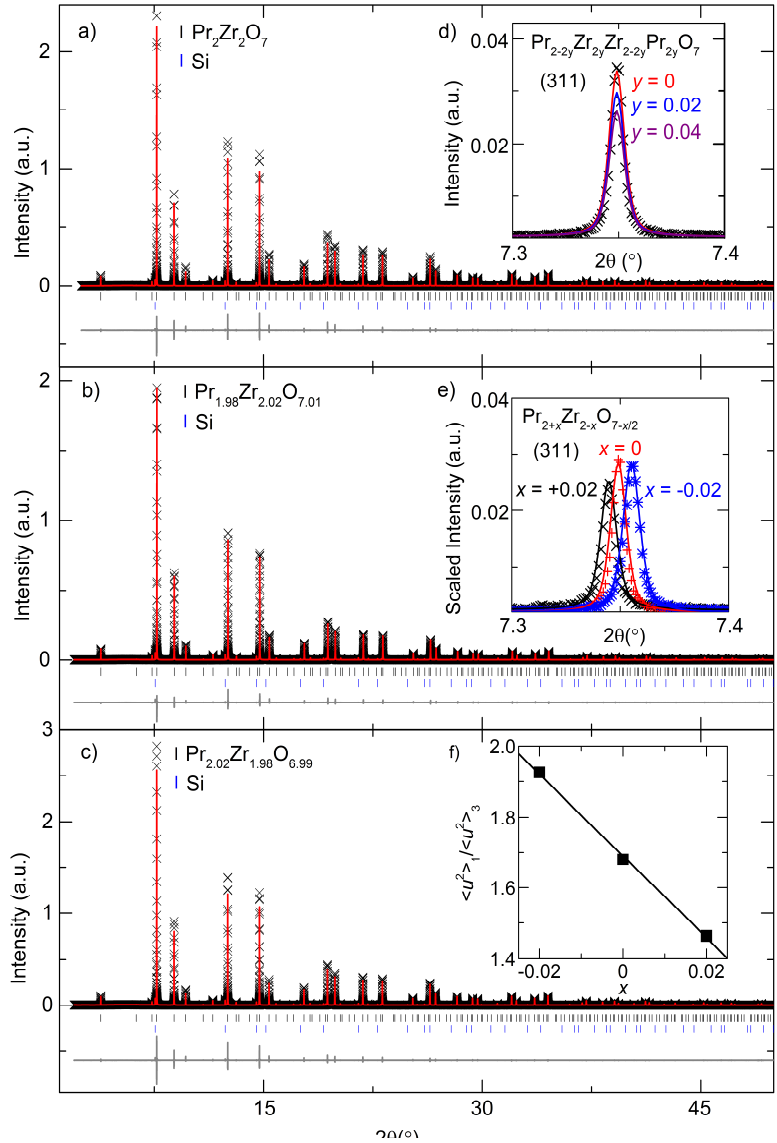

**Fig. 3.** Rietveld refinement of synchrotron X-ray diffraction data of $Pr_{2+x}Zr_{2-x}O_{7-x/2}$ for (a) $x = 0$, (b) $x = -0.02$, and (c) $x = +0.02$. To the precision of these data (~1%) samples have the ideal stoichiometry and no A/B site mixing. (d) As an example of our sensitivity to Pr/Zr site mixing, the change in model intensity for the (311) pyrochlore-only reflection for various mixing ratios is compared to the observed data for $x = 0$. (e) The changes in the (311) pyrochlore reflection with composition is accurately modeled by the changes in Pr/Zr ratio, without any additional Pr/Zr mixing that would indicate a transition to a fluorite-like structure. (f) The ratio of mean square displacements ($<u^2>$) of Pr as $x$ changes, with a guide to the eye. This ratio allocates a numerical value to the pancake-like shape of the Pr ellipsoids ($<u^2>_1 = <u^2>_2 > <u^2>_3$) which are more pancake-like in the $x = -0.02$ specimen and suggestive of a local off-centering of the Pr ions, consistent with recent neutron scattering measurements [46].

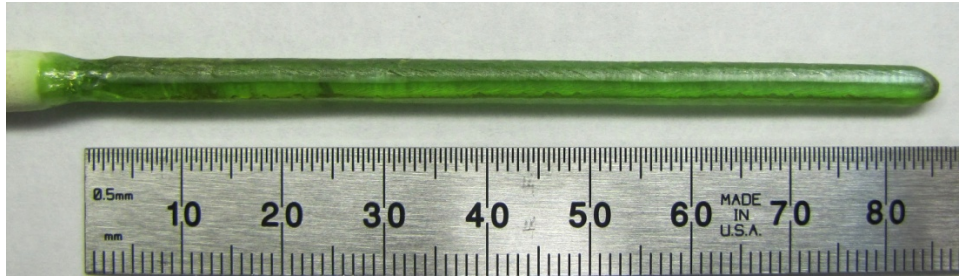

**Fig. 4.** A typical $Pr_2Zr_2O_7$ single crystal grown under 1 bar static high-purity argon atmosphere using an optical floating zone image furnace.

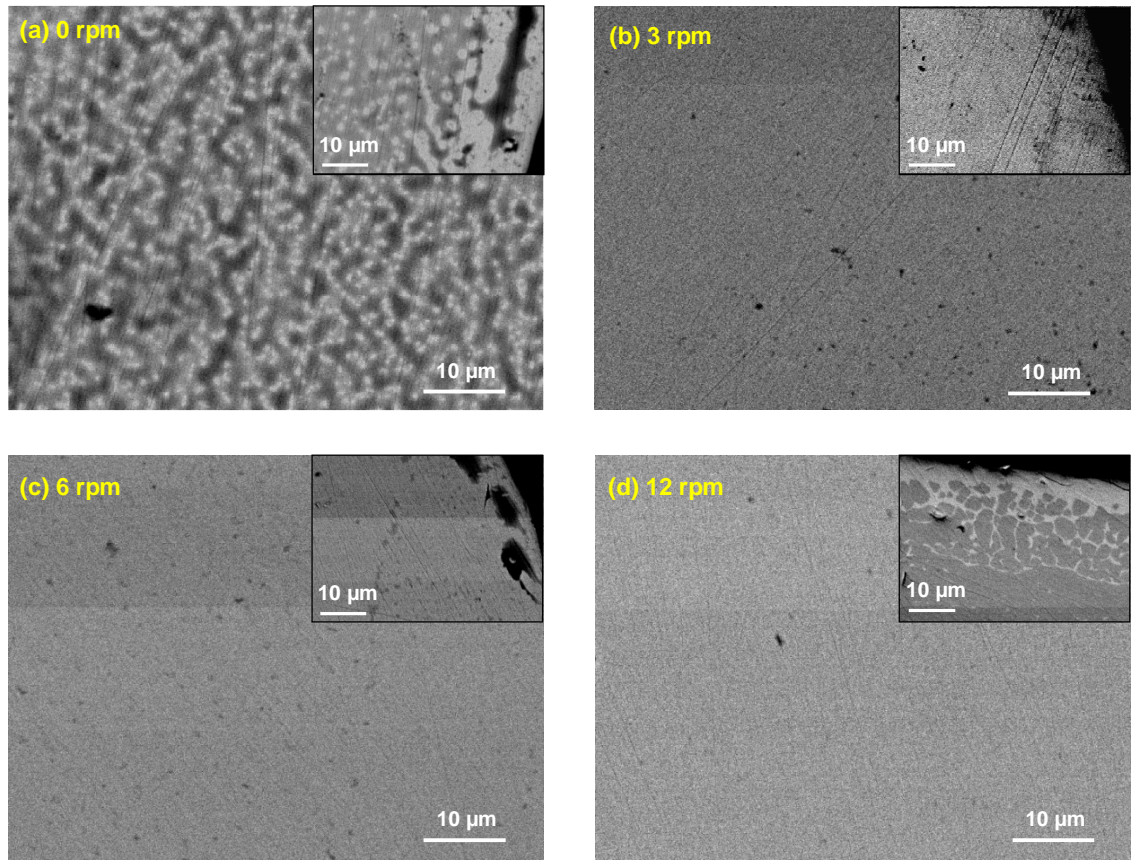

**Fig. 5.** SEM back scattered micrographs of cross sections, taken from inside (large pictures) and close to the edge near the surface (smaller pictures on the top right), cut from $Pr_2Zr_2O_7$ single crystals grown at (a) 0 rpm, (b) 3 rpm, (c) 6 rpm and (d) 12 rpm. A spinodal decomposition is shown for the growth at 0 rpm, while applying the rotation to 3, 6 and 12 rpm forms a more uniform and homogeneous internal microstructure, with the highest rotation rates driving excess Pr to the edges producing inhomogeneity at the surface.

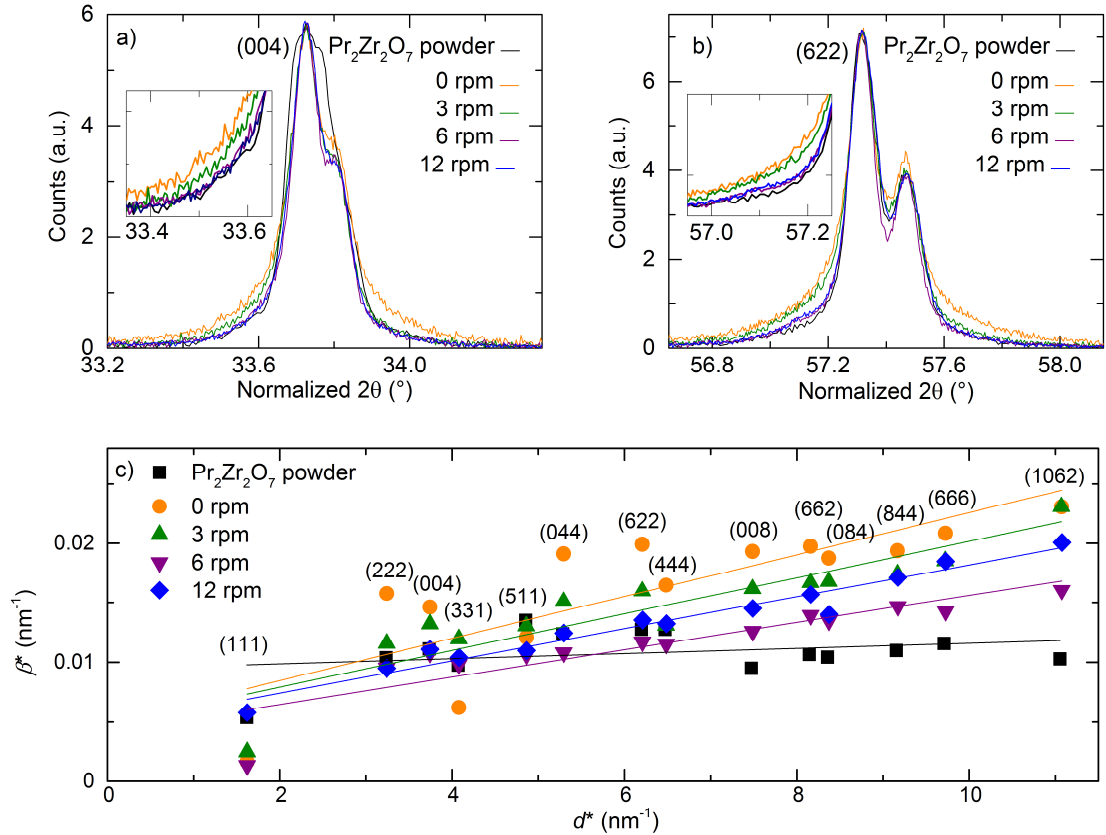

**Fig. 6.** Representative (a) (004) and (b) (622) peaks of ground, single crystalline $Pr_2Zr_2O_7$ samples grown with different rotation rates as well as a stoichiometric powder sample. The peak intensities and positions are normalized to properly visualize the peak broadening. (c) A Williamson-Hall plot with linear fits of the form $\beta^* = 0.5\varepsilon d^* + K/L$, with the peaks labeled appropriately. Here $\beta^* = \beta_{obs}\cos\theta\,\lambda^{-1}$, $d^* = 2\sin\theta\,\lambda^{-1}$, $\varepsilon$ is the apparent strain, K is a constant close to unity, and L is the average crystallite size. The strain is minimized for a rotation rate of 6 rpm.

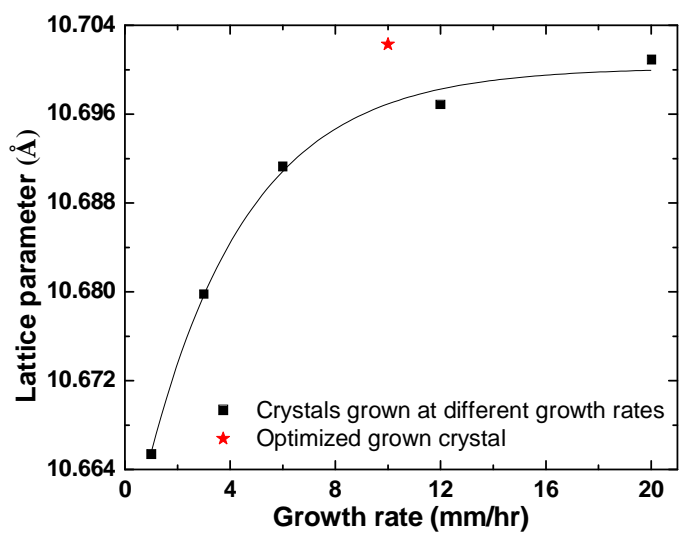

**Fig. 7.** Lattice parameter measurements of crystals grown at different growth rates of 1-20 mm/hr, shown in solid black squares (solid line is guide to the eye). Increasing the growth speed resulted in the vaporization of less Pr, and therefore, obtain crystals with larger lattice parameters close to the stoichiometric powder sample ($a$ = 10.70254(4) Å). For comparison, the lattice parameter of a crystal grown under optimized conditions of 10 mm/hr with an extra 1 atomic % Pr added into the starting feed rods to offset Pr vaporization is shown.

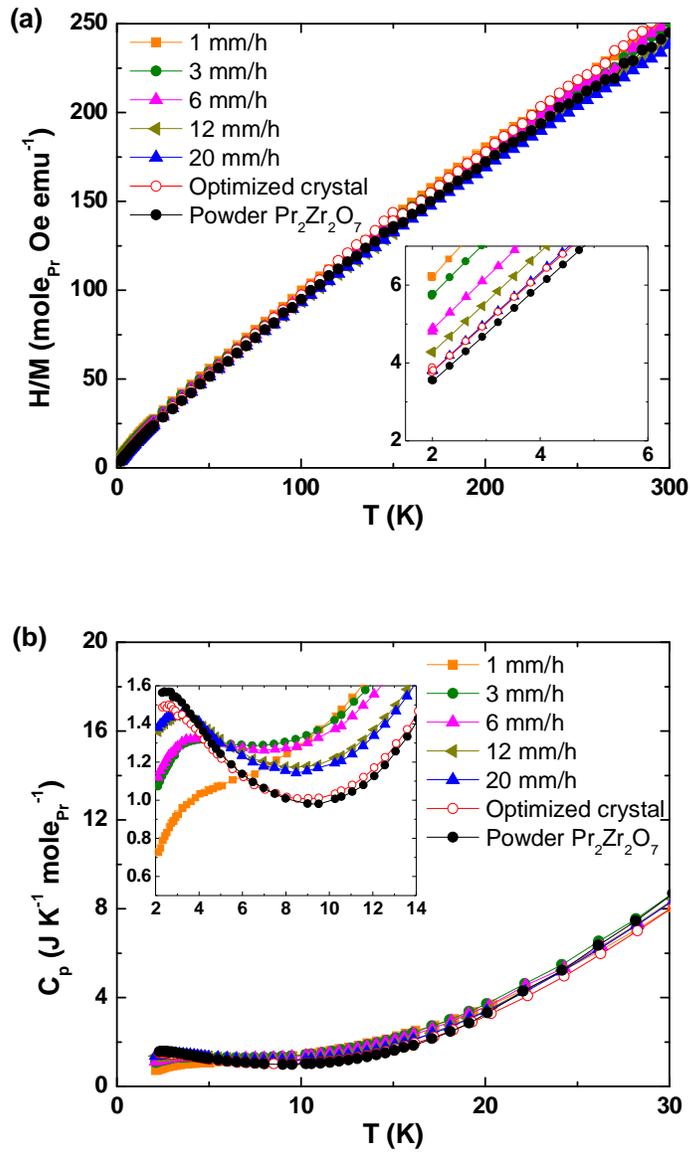

**Fig. 8.** Temperature dependence of inverse magnetic susceptibility (a) and specific heat (b) of single crystalline samples grown at different growth rates. Insets are shown for a closer look at low temperatures. A crystal grown under optimized conditions exhibits the same physical properties as the stoichiometric powder sample, demonstrating its high quality.

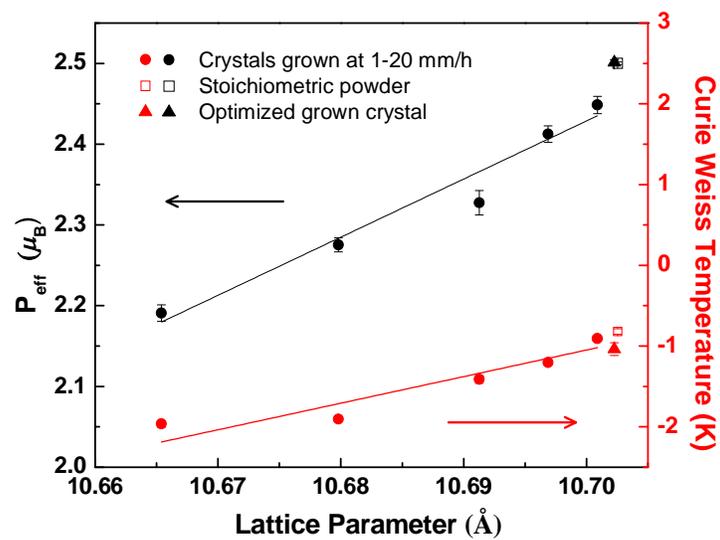

**Fig. 9.** Curie-Weiss fitting results for the data in Fig. 8a. The solid lines are linear fits. Results from a crystal grown under optimized conditions are within error of those for the stoichiometric powder reference.

**Table I.** Structural Parameters for $Pr_{2+x}Zr_{2-x}O_7$ using $Fd$-$3m$:2 (227, origin 2), obtained from Rietveld refinements of room temperature synchrotron powder diffraction data. Atomic positions are restricted by symmetry as Pr: 16$d$ (1/2, 1/2, 1/2), Zr: 16$c$ (0 0 0), O: 48$f$ ($x$, 1/8, 1/8), O': 8$b$ (3/8, 3/8, 3/8).

| | Parameter | $Pr_{2.02}Zr_{1.98}O_{6.99}$ | $Pr_2Zr_2O_7$ | $Pr_{1.98}Zr_{2.02}O_{7.01}$ |
|---|---|---|---|---|
| | $a$ (Å) | 10.710322(3) | 10.703798(3) | 10.694711(3) |
| | Cell Volume | 1228.5918(6) | 1226.3480(6) | 1223.2272(6) |
| Pr | occ* (Pr/Zr) | 1.0/0 | 1.0/0 | 0.99/0.01 |
| | $U_{eq}$ (Å$^2$) | 0.0083 | 0.0078 | 0.0084 |
| | $U_{11}=U_{22}=U_{33}$ (Å$^2$) | 0.00825(3) | 0.00776(3) | 0.00844(2) |
| | $U_{12}=U_{23}=U_{13}$ (Å$^2$) | -0.00109(10) | -0.00121(9) | -0.00161(7) |
| Zr | occ* (Zr/Pr) | 0.99/0.01 | 1.0/0 | 1.0/0 |
| | $U_{eq}$ (Å$^2$) | 0.0044 | 0.0043 | 0.0054 |
| | $U_{11}=U_{22}=U_{33}$ (Å$^2$) | 0.00440(4) | 0.00428(4) | 0.00559(3) |
| | $U_{12}=U_{23}=U_{13}$ (Å$^2$) | 0.00003(14) | 0.00043(13) | 0.00077(10) |
| O | $x$ | 0.33368(15) | 0.33391(15) | 0.33428(10) |
| | occ* | 1.0 | 1.0 | 1.0 |
| | $U_{iso}$ (Å$^2$) | 0.0076(4) | 0.0080(4) | 0.0081(3) |
| O' | occ* | 1.0 | 1.0 | 1.0 |
| | $U_{iso}$ (Å$^2$) | 0.0113(10) | 0.0085(9) | 0.0069(6) |
| | $\chi^2$ | 4.071 | 6.802 | 3.285 |
| | $R_{wp}$ (%) | 12.08 | 12.01 | 8.69 |
| | $R_p$ (%) | 9.83 | 9.68 | 7.31 |
| | $R_F^2$ (%) | 4.96 | 5.15 | 3.47 |

*Occupancies were fixed at nominal values for final refinement.

**Table II.** Williamson-Hall Fit Parameters

| Rotation Rate | $\varepsilon$ | L/K (nm) |
|---|---|---|
| Powder $Pr_2Zr_2O_7$ | 0.0001(1) | 106(15) |
| 0 rpm | 0.00088(19) | 200(100) |
| 3 rpm | 0.00077(11) | 200(60) |
| 6 rpm | 0.00057(9) | 240(70) |
| 12 rpm | 0.00067(4) | 210(30) |